\documentclass[prl,twocolumn,showpacs]{revtex4}
\usepackage{amsmath,amssymb,mathrsfs}
\usepackage{psfrag}
\usepackage{graphicx}
\usepackage{graphics}
\usepackage{epsfig}
\usepackage{bm}
\usepackage{color}
\usepackage{verbatim,color,ulem}

\newcommand{\beq}{\begin{equation}}
\newcommand{\eeq}{\end{equation}}
\newcommand{\beqa}{\begin{eqnarray}}
\newcommand{\eeqa}{\end{eqnarray}}

\begin{document}

\title{Composite bosons in the 2D BCS-BEC crossover 
from Gaussian fluctuations}
\author{L. Salasnich and F. Toigo}
\affiliation{
Dipartimento di Fisica e Astronomia ``Galileo Galilei'' 
and CNISM, Universit\`a di Padova, Via Marzolo 8, 35131 Padova, Italy}

\date{\today}

\begin{abstract}
We study Gaussian fluctuations of the zero-temperature attractive Fermi gas 
in the 2D BCS-BEC crossover showing that they are crucial to get a 
reliable equation of state in the BEC regime of composite bosons, bound 
states of fermionic pairs. 
A low-momentum expansion up to the fourth order of the quadratic 
action of the fluctuating pairing field gives an ultraviolent divergent 
contribution of the Gaussian fluctuations to the grand potential. 
Performing dimensional regularization we evaluate the effective 
coupling constant  in the beyond-mean-field grand potential. Remarkably, in the 
BEC regime our grand potential gives exactly the Popov's equation 
of state of 2D interacting bosons, and allows us to identify 
the scattering length $a_B$ of the interaction between composite bosons as
$a_B=a_F/(2^{1/2}e^{1/4})= 0.551... a_F$, with $a_F$ is the scattering 
length of fermions. Remarkably, the value from our analytical relationship 
between the two scattering lengths is in full agreement with that obtained by
recent Monte Carlo calculations. 
\end{abstract}


\maketitle


Thermal and quantum fluctuations play a relevant role 
in any generic 2D superfluid system 
\cite{mermin,hohenberg,coleman,nagaosa}. 
Triggered by the experimental realization of the BCS-BEC crossover 
with three-dimensional ultracold atoms 
\cite{exp_MolecularBEC,exp_Crossover,chin}, 
in the last years several theoretical papers 
\cite{loktev0,babaev,loktev,demelo,tempere,tempere2,sala} 
have been devoted to the study of thermal fluctuations 
in the two-dimensional (2D) BCS-BEC crossover, 
i.e in the crossover of a 2D fermionic superfluid 
from weakly-bound BCS-like Cooper pairs to the Bose-Einstein condensation 
(BEC) of strongly-bound molecules. Recently, 
zero-temperature quantum effects beyond the old mean-field predictions 
of Randeria, Duan and Shieh \cite{randeria2d} 
have been investigated by Bertaina and Giorgini 
\cite{bertaina}. By using the fixed-node diffusion 
Monte Carlo (MC) numerical method they have found that 
in the BEC regime the zero-temperature MC equation of state 
shows dimer-dimer and atom-dimer interaction effects 
that are completely neglected in the mean-field picture \cite{bertaina}. 

In this Rapid Communication we study quantum fluctuations  
of the zero-temperature attractive Fermi gas 
in the 2D BCS-BEC crossover by using a path-integral approach. 
Our theoretical analysis is the 2D counterpart of similar 
beyond-mean-field investigations performed in the 3D 
crossover \cite{pieri,hu,randeria3D}. 
However, the calculations presented in the above mentioned references  
are by no means readily extended from 3D to 2D. In fact, 
at our knowledge, no extension has been made to 2D of the regularization 
approach of Refs. \cite{pieri,hu,randeria3D} based on convergence factors. 
For this reason we adopt a completely different method, 
i.e. the dimensional regularization plus flow equation 
quantum-field-theory technique, which is widely 
used in high-energy particle physics, but rarely encountered 
in condensed matter theory

We show that, contrary to the 3D case, 
in our 2D fermionic system 
the interaction between composite bosons is fully induced by 
quantum fluctuations. To obtain this intriguing result we investigate 
the zero-point energy of collective 
bosonic excitations obtained from the the quadratic Gaussian  
action of the fluctuating pairing field. 
This divergent zero-point energy can be set to zero 
on the basis of dimensional regularization \cite{dim-reg} 
only if one considers a low-momentum expansion 
of the Gaussian action up to second order \cite{leibb}. Here we perform 
a low-momentum expansion up to the fourth order and 
using dimensional regularization we find a running coupling 
constant from which we 
derive an effective beyond-mean-field grand potential. Remarkably, 
from this effective grand potential, which includes quantum fluctuations,  
we find exactly the recursive 
Popov's equation of state \cite{popov} of 2D interacting bosons 
with its beyond-mean-field logarithmic correction, which reduces 
to the Schick's equation of state \cite{schick} at the 
leading order \cite{boronat}. In particular, 
we find that the scattering length $a_B$ of composite 
bosons is given by $a_B=a_F/(2^{1/2}e^{1/4})= 0.551... a_F$, 
with $a_F$ is the scattering length of fermions. 
This fully-analytical result is in very good agreement with recent Monte Carlo 
calculations of Bertaina and Giorgini \cite{bertaina} and previous 
four-body scattering calculations of Petrov, Baranov 
and Shlyapnikov \cite{baranov}. 


{\it The model}. We consider a two-dimensional 
attractive Fermi gas of ultracold and dilute 
two-spin-component neutral atoms. We adopt the path integral formalism, 
where the atomic fermions are described by the complex Grassmann 
fields $\psi_{\sigma} ({\bf r},\tau )$, $ \bar{\psi}_{\sigma} ({\bf r},\tau )$ 
with spin $\sigma = ( \uparrow , \downarrow )$  \cite{nagaosa}. 
The Euclidean Lagrangian density of the uniform 
system in a two-dimensional box of area $L^2$ 
and with chemical potential $\mu$ is given by 
\beq 
\mathscr{L} = \bar{\psi}_{\sigma} \left[ \hbar \partial_{\tau} 
- \frac{\hbar^2}{2m}\nabla^2 - \mu \right] \psi_{\sigma} 
+ g \, \bar{\psi}_{\uparrow} \, \bar{\psi}_{\downarrow} 
\, \psi_{\downarrow} \, \psi_{\uparrow} \; , 
\label{lagrangian-initial}
\eeq
where $g<0$ is the strength of the s-wave inter-atomic 
coupling \cite{nagaosa}. 
Summation over the repeated index $\sigma$ in the Lagrangian is meant. 
The interaction strength $g$ of s-wave pairing is related to the 
binding energy $\epsilon_b$ of a fermion pair in  vacuum by 
the expression \cite{randeria2d,marini}
\beq 
- \frac{1}{g} = \frac{1}{2L^2} \sum_{\bf k} \frac{1}{\epsilon_k + 
\frac{1}{2} \epsilon_b} \; . 
\label{g-eb}
\eeq
Note that, contrary to the 3D case, in 2D realistic interatomic potentials 
have always a bound state \cite{bertaina,marini}. In addition, 
according to Mora and Castin \cite{castin1} 
the binding energy $\epsilon_b$ of two fermions can be written 
in terms of the 2D fermionic scattering length $a_F$ as 
\beq 
\epsilon_b= {4\over e^{2\gamma}}{\hbar^2\over m a_F^2} \; ,   
\label{eb-af}
\eeq
where $\gamma=0.577...$ is the Euler-Mascheroni constant.
 
Through the usual Hubbard-Stratonovich transformation 
\cite{nagaosa} the Lagrangian density $\mathscr{L}$ of 
Eq. (\ref{lagrangian-initial}), 
quartic in the fermionic fields, 
can be rewritten as a quadratic form by introducing the
auxiliary complex scalar field $\Delta({\bf r},\tau)$ so that 
\beq 
\mathscr{L}_e =
\bar{\psi}_{\sigma} \left[  \hbar \partial_{\tau} 
- {\hbar^2\over 2m}\nabla^2 - \mu \right] \psi_{\sigma} 
+ \bar{\Delta} \, \psi_{\downarrow} \, \psi_{\uparrow} 
+ \Delta \bar{\psi}_{\uparrow} \, \bar{\psi}_{\downarrow} 
- {|\Delta|^2\over g} \; . 
\label{ltilde}
\eeq 

The partition function ${\cal Z}$ of the 
system at temperature $T$ can then be written as 
\beq 
{\cal Z} = \int {\cal D}[\psi_{\sigma},\bar{\psi}_{\sigma}]\, 
{\cal D}[\Delta,\bar{\Delta}] \ 
\exp{\left\{ - {S_e(\psi_{\sigma}, \bar{\psi_{\sigma}},
\Delta,\bar{\Delta}) \over \hbar} \right\}} \; , 
\label{papo}
\eeq
where 
\beq 
S_e(\psi_{\sigma}, \bar{\psi_{\sigma}},\Delta,\bar{\Delta}) 
= \int_0^{\hbar\beta} 
d\tau \int_{{L^2}} d^2{\bf r} \ 
\mathscr{L}_e(\psi_{\sigma}, \bar{\psi_{\sigma}},\Delta,\bar{\Delta})
\label{e-action}
\eeq
is the effective action and $\beta \equiv 1/(k_B T)$ with $k_B$ 
Boltzmann's constant.

{\it Review of mean-field results}. We shall investigate the effect of 
fluctuations of the gap field $\Delta({\bf r},t)$ around its
mean-field value $\Delta_0$ which may be taken to be real. 
For this reason we set 
\beq 
\Delta({\bf r},\tau) = \Delta_0 +\eta({\bf r},\tau)  \; , 
\label{polar}
\eeq
where $\eta({\bf r},\tau)$ is the complex paring field of bosonic 
fluctuations \cite{nagaosa}. 

Mean-field results are obtained neglecting bosonic fluctuations, 
i.e. setting $\eta({\bf r},t)=0$.  
Integrating over the fermionic fields  $\psi_s({\bf r},t)$ 
and $\bar{\psi}_s({\bf r},t)$ in Eq. (\ref{papo}) 
one finds immediately the mean-field partition 
function \cite{nagaosa,loktev0,babaev,loktev,sala}
\beq 
{\cal Z}_{mf} =  \exp{\left\{ - {S_{mf}\over \hbar} \right\}}
= \exp{\left\{ - \beta \, \Omega_{mf} \right\}} \; , 
\eeq
where 
\beqa 
{S_{mf}\over \hbar} &=& - Tr[\ln{(G_0^{-1})}] - 
\beta {L^2} {\Delta_0^2\over g} \; 
\nonumber
\\
&=& - \sum_{{\bf k}} \left[ 2
\ln{\left( 2 \cosh{(\beta E_{sp}(k)/2)} \right)} 
- \beta (\epsilon_k -\mu) \right] 
\nonumber
\\
&-& \beta L^2 {\Delta_0^2\over g} \; , 
\label{omega-sp} 
\eeqa
with $\epsilon_k=\hbar^2k^2/(2m)$, 
\beq 
G_0^{-1} = \left(
\begin{array}{cc}
\hbar \partial_{\tau} -{\hbar^2\over 2m}\nabla^2 -\mu & \Delta_0 \\ 
\Delta_0 & \hbar \partial_{\tau} +{\hbar^2\over 2m}\nabla^2 +\mu
\end{array}
\right)
\label{G0}
\eeq
the inverse mean-field Green function, and 
\beq 
E_{sp}(k)=\sqrt{(\epsilon_k-\mu)^2+\Delta_0^2} 
\label{ex-fermionic}
\eeq 
the energy of the fermionic single-particle elementary excitations. 

At zero temperature ($T=0$, i.e. $\beta\to +\infty$) 
the mean-field grand potential $\Omega_{mf}$ becomes 
\beq 
\Omega_{mf} = - \sum_{\bf k} \left( E_{sp}(k) - \epsilon_k + \mu \right) 
- L^2 {\Delta_0^2\over g} \; . 
\label{omega0-div}
\eeq
In the continuum limit $\sum_{\bf k}\to L^2\int d^2{\bf k}/(2\pi)^2$ 
the logarithmic divergence of the grand potential $\Omega_{mf}$ 
is removed by using Eq. (\ref{g-eb}), 
which gives the interaction strength $g$ in terms of the binding 
energy $\epsilon_b$ of pairs. In this way one obtains 
\beqa 
\Omega_{mf} = - {m L^2\over 4\pi \hbar^2} \Big[ 
\mu^2 + \mu \sqrt{\mu^2+\Delta_0^2} + {1\over 2} \Delta_0^2 
\nonumber
\\
- \Delta_0^2 \ln{\Big({-\mu + \sqrt{\mu^2 + \Delta_0^2} 
\over \epsilon_b}\Big)} \Big]  \; . 
\label{omega-mf-full}
\eeqa
The constant, uniform and real gap parameter $\Delta_0$ is obtained 
by minimizing $\Omega_{mf}$ with respect to $\Delta_0$, namely
\beq 
\left({\partial \Omega_{mf}\over \partial \Delta_0}\right)_{\mu,L^2} = 0 \; ,  
\eeq 
from which one finds the gap equation 
\beq
\Delta_0 = \sqrt{2\epsilon_b ( \mu + {1\over 2}\epsilon_b ) } \; , 
\label{gapeq}
\eeq 
which gives the energy gap $\Delta_0$ as a function of 
the chemical potential $\mu$ and the binding energy $\epsilon_b$. Inserting 
this formula into Eq. (\ref{omega-mf-full}) we find 
\beq 
\Omega_{mf} = - {m L^2\over 2\pi \hbar^2} 
(\mu + {1\over 2} \epsilon_b )^2  \; . 
\label{omega-mf}
\eeq
The total number density $n=N/L^2$ of fermions is obtained from the
familiar zero-temperature thermodynamic relation
\beq
n = - {1\over L^2} {\partial \Omega_{mf}\over \partial \mu}
\eeq
which immediately gives the chemical potential $\mu$ as a 
function of the number density $n=N/L^2$, i.e. 
\beq 
\mu = {\pi \hbar^2 \over m} n - {1\over 2} \epsilon_b \; . 
\label{echem-mf}
\eeq
This is the mean-field equation of state of the 2D superfluid Fermi gas 
in the BCS-BEC crossover obtained by 
Randeria, Duan and Shieh \cite{randeria2d}. 
In the BCS regime, where $\epsilon_b \ll \epsilon_F$ 
with $\epsilon_F=\pi\hbar^2n/m$ the Fermi energy of the 2D ideal Fermi gas, 
one finds $\mu \simeq \epsilon_F >0$ while in the BEC regime, 
where $\epsilon_b \gg \epsilon_F$ one has $\mu \simeq - \epsilon_b/2 <0$. 

Introducing $\mu_B = 2(\mu + \epsilon_b/2)$ as the chemical potential 
of composite bosons (made of bound fermionic pairs) with mass $m_B=2m$ and 
density $n_B=n/2$,  we may rewrite the above equation of state 
Eq. (\ref{echem-mf})in terms of bosonic quantities as: 
\beq 
\mu_B  = {8\pi \hbar^2 \over m_B} n_B \; . 
\eeq

Clearly, this mean-field equation of state showing a
bosonic chemical potential $\mu_B$  independent of the interaction 
between bosons is lacking important informations which must be 
encoded in quantum fluctuations. 
As previously explained, the main goal of this paper 
is to take into account these quantum fluctuations, which are 
crucial in reduced dimensionalities\cite{mermin,hohenberg,coleman,nagaosa}. 


{\it Gaussian quantum fluctuations}. 
We now consider the effect of quantum fluctuations, i.e. in Eq. (\ref{polar})
we allow $\eta({\bf r},t)\neq 0$. 
Expanding the effective action $S_e(\psi_s, \bar{\psi_s},\Delta,\bar{\Delta})$ 
of Eq. (\ref{e-action}) around $\Delta_0$ 
up to the quadratic (Gaussian) order in $\eta({\bf r},t)$ 
and $\bar{\eta}({\bf r},t)$ one finds 
\beq 
Z = Z_{mf} \ \int 
{\cal D}[\eta,\bar{\eta}] \ 
\exp{\left\{ - {S_g(\eta,\bar{\eta}) \over \hbar} \right\}} \; , 
\label{sigo}
\eeq
where 
\beq 
S_{g}(\eta,\bar{\eta}) = {1\over 2} \sum_{Q} 
({\bar\eta}(Q),\eta(-Q)) \ {\bf M}(Q) \left(
\begin{array}{c}
\eta(Q) \\ 
{\bar\eta}(-Q) 
\end{array}
\right) \; 
\eeq
is the Gaussian action of fluctuations in the reciprocal space 
with $Q=({\bf q},i\nu_m)$ 
the $3$-vector denoting the momenta ${\bf q}$ and Matsubara 
frequencies $\nu_m=2\pi m/\beta$. 
Integrating over the bosonic fields $\eta(Q)$ 
and $\bar{\eta}(Q)$ in Eq. (\ref{sigo}) one finds 
the Gaussian grand potential \cite{randeria3D,schakel,tempere2,marini2}
\beq 
\Omega_g = {1\over 2\beta} \sum_{Q} \ln{\mbox{Det}({\bf M}(Q))}
\eeq
The $2\times 2$ matrix ${\bf M}(Q)$ 
is the inverse fluctuation propagator, whose non trivial dependence 
on $Q$ can be found in Refs. \cite{tempere2,marini2}. 
$\mbox{Det}({\bf M}(Q))=\mbox{Det}({\bf M}({\bf q},z))$ has zero on the real 
axis of the $z$ complex plane at $z=\pm \omega_0({\bf q})$ which 
correspond to the poles of the fluctuation propagator, and 
describe the spectrum $E_{col}(q)=\hbar \, \omega_0(q)$ of the bosonic 
collective excitations \cite{randeria3D,schakel,tempere2,marini2}. 
These excitations can be extracted from ${\bf M}(Q)$ with a 
low-energy and low-momentum expansion \cite{tempere2,marini2} 
to give:
\beq 
E_{col}(q) = 
\sqrt{\epsilon_q \left( \lambda \ \epsilon_q + 2 \ m \ c_s^2 \right)} 
\label{spettrob}
\eeq
where $\epsilon_q=\hbar^2q^2/(2m)$ is the free-particle energy, 
$\lambda$ takes into account the first correction 
to the familiar low-momentum phonon dispersion 
$E_{col}(q) \simeq c_s \hbar q$, with $c_s$ is the sound velocity. 
Both $\lambda$ and $c_s$ depend on the chemical potential $\mu$ and 
the energy gap $\Delta_0$, which is itself a function of $\mu$ 
and $\epsilon_b$ on the basis of the gap equation (\ref{gapeq}). 
In particular, one finds \cite{marini,marini2}
\beq 
\lambda = {4x_0^2+1-8x_0\sqrt{x_0^2+1}
\over 
24 \sqrt{x_0^2+1} (\sqrt{x_0^2+1}-x_0)}
\label{lambda}
\eeq
with $x_0=\mu/\Delta_0=((\mu+\epsilon_b/2)-\epsilon_b/2)/
\sqrt{2\epsilon_b(\mu+\epsilon_b/2)}$ and 
\beq 
m \ c_s^2 = {\Delta_0\over 2} \left( x_0 + \sqrt{x_0^2+1} \right) = 
\mu + {1\over 2} \epsilon_b \; ,  
\label{cs}
\eeq
where the last equality is obtained using Eq. (\ref{gapeq}). 
An inspection of Eq. (\ref{lambda}) shows that $\lambda$ is positive 
if $x_0 <0.132$ and it goes quickly to $\lambda =1/4$ in the BEC regime, 
where $x_0$ is large and negative, i.e. for $-x_0\gg 1 $. Thus in the 
BEC regime the spectrum (\ref{spettrob}) of collective bosonic 
excitations reduces to the familiar Bogoliubov 
spectrum \cite{nagaosa,schakel} of bosonic excitations with mass $m_B=2m$. 
Instead, $\lambda$ is negative for $x_0>0.132$ and it goes to 
$\lambda=-x_0^2/3$ for $x_0\gg 1$ that is the BCS regime. 
At zero temperature, the total grand potential finally reads 
\beq 
\Omega = - \lim_{\beta \to +\infty} {1\over \beta} \ln{(Z)} 
= \Omega_{mf} + \Omega_{g} \; 
\eeq
where $\Omega_{mf}$ is given by Eq. (\ref{omega-mf}), 
while $\Omega_{g}$ reads
\beq 
\Omega_{g} = {1\over 2} \sum_{{\bf q}} E_{col}(q) \; . 
\label{omegacol-div}
\eeq
This is the zero-point energy of bosonic collective excitations, 
i.e. the zero-temperature Gaussian fluctuations. 
In the continuum limit Eq. (\ref{omegacol-div}) is ultraviolet divergent 
if $\lambda>0$. Instead, if $\lambda <0$ 
the spectrum (\ref{spettrob}) has a natural ultraviolet 
cutoff $q_c$, given by $\hbar^2 q_c^2/(2m) = 2 mc_s^2/|\lambda|$, which 
goes to zero in the deep BCS regime where, consequently, quantum 
fluctuations are strongly suppressed. 


{\it Analysis in the BEC regime}. 
Since we are interested in the BEC regime ($\lambda>0$ and 
in particular $\lambda=1/4$) we must regularize Eq. (\ref{omegacol-div}). 
To this end we use the dimensional regularization \cite{dim-reg,schakel}, 
i.e. we extend the two-dimensional integral to a generic complex 
$D=2-\varepsilon$ dimension, and then take the limit $\varepsilon \to 0$. 
In this way 
\beqa 
{\Omega_{g} \over L^D} &=& {1\over 2} 
\int {d^{D}{\bf q} \over (2\pi)^{D}} E_{col}(q) 
\nonumber 
\\
&=& - {A(0) \over 2 \kappa^{\varepsilon}} (mc_s^2)^2 \ 
\Gamma(-2+{1\over 2}\varepsilon) \; ,
\label{om_eps} 
\eeqa
where the regulator $\kappa$ is an arbitrary scale wavenumber 
which enters for dimensional reasons. 
In Eq. $(\ref{om_eps})$ we have defined 
$A(0)=m/(2\pi\hbar^2 \lambda^{3/2})$ and 
$\Gamma(z)$ is the Euler gamma function, such that 
$\Gamma(-2+\varepsilon/2)=1/\varepsilon + O(\varepsilon^0)$ 
for $\varepsilon\to 0$. Consequently, 
using Eq. (\ref{cs}), to leading order in $1/\varepsilon$ 
we get \cite{schakel0}
\beq 
{\Omega_{g} \over L^D} = - 
{A(0)\over 2\varepsilon \, \kappa^{\varepsilon}} 
(\mu + {1\over 2}\epsilon_b)^2 \; . 
\label{biro}
\eeq
This expression is still divergent. Nevertheless, comparing $\Omega_g$ 
with $\Omega_{mf}$ in $D=2-\varepsilon$ dimensions 
(see also Eq. (\ref{omega-mf})) given by   
\beq 
{\Omega_{mf} \over L^D} = - {1\over 2 \xi(\varepsilon)} 
(\mu + {1\over 2} \epsilon_b)^2  
\eeq 
with $\xi(\varepsilon)
=(4\pi\hbar^2/m)^{1-\varepsilon/2}\epsilon_b^{\varepsilon/2}
(1+\varepsilon/2)/(4\Gamma(1+\varepsilon/2))$ 
the mean-field coupling constant \cite{remark}, 
we conclude that the total grand potential reads 
\beq 
{\Omega\over L^2} = {\Omega_{mf}\over L^2} + 
{\Omega_g \over L^2} = - {1\over 2 \xi_r(k,\varepsilon)} 
(\mu + {1\over 2} \epsilon_b)^2 \; , 
\label{totale}
\eeq 
where it appears the renormalized coupling 
constant $\xi_r(\kappa,\varepsilon)$ given by 
\beq 
{1\over \xi_r(\kappa,\epsilon)} 
= \kappa^{\varepsilon} \left( {1\over \xi(\varepsilon)} + 
{A(0 ) \over \varepsilon \, \kappa^{\varepsilon}} \right) \; . 
\label{rodi}
\eeq
The parameter $\xi_r(\kappa,\epsilon)$ 
is the ``running coupling constant'' of our 
theory which runs by changing $\kappa$ \cite{schakel,kaku}. 
To extract its dependence on $\kappa$ 
we introduce the flow  function 
$\beta(\xi_r)\equiv \kappa d\xi_r/d\kappa$, which encodes 
the dependence of the renormalized coupling 
constant $\xi_r$ on the wavenumber scale $\kappa$ \cite{kaku}
\beq 
\beta(\xi_r)=\kappa {d\xi_r \over d\kappa} = 
A(0) \ \xi_r^2 + O(\varepsilon) \; . 
\label{deq}
\eeq

After integration of Eq. (\ref{deq}) in the limit 
$\varepsilon \to 0$ we get 
\beq 
{1\over \xi_r(\kappa',0)} - {1\over \xi_r(\kappa,0)} 
= - A(0) \ \ln{\left({\kappa'\over \kappa}\right)} \; , 
\label{simsalabim}
\eeq
where  $A(0)=m/(2\pi\hbar^2 \lambda^{3/2})$. 
We set the Landau pole \cite{kaku} 
of Eq. (\ref{simsalabim}) at the high energy scale of the system 
$\epsilon_b$, i.e. we set  $1/\xi_r(\kappa',0)=0$ at $\kappa'$ such that  
$\hbar^2\kappa'^2/(2m)=\epsilon_b/2$. Then, when $\kappa$ corresponds 
to the actual energy of our system, i.e. 
$\hbar^2\kappa^2/(2m)=\mu+\epsilon_b/2$, 
from Eqs. (\ref{totale}) 
with $\varepsilon \to 0$ we obtain 
\beq 
\Omega = - {m L^2\over 8\pi\hbar^2 \lambda^{3/2}}  
(\mu + {1\over 2}\epsilon_b)^2 \ \ln{\left({\epsilon_b\over 
2 (\mu + {1\over 2}\epsilon_b) } \right)} \; .  
\eeq

In this BEC limit, where $\lambda=1/4$, introducing again 
$\mu_B = 2(\mu + \epsilon_b/2)$ as the chemical potential 
of composite bosons with mass $m_B=2m$ and 
density $n_B=n/2$, the total grand potential can be rewritten as 
\beq 
\Omega = - {m_B L^2\over 8\pi\hbar^2} 
\mu_B^2 \ \ln{\left({\epsilon_b\over \mu_B } \right)} \; . 
\eeq
As usual, the total density of bosons $n_B=n/2$ is obtained 
in terms of $\mu_B=2(\mu+\epsilon_b/2) $ from the zero-temperature 
thermodynamic formula 
\beq 
n = - {1\over L^2} {\partial \Omega\over \partial \mu} \; , 
\eeq 
which leads to: 
\beq 
n_B = {m_B\over 4\pi \hbar^2} \mu_B 
\ln{\left({\epsilon_b \over \mu_B \ e^{1/2}} \right)} \; . 
\label{law}
\eeq
 
Inserting Eq. (\ref{eb-af}), which gives the binding energy $\epsilon_b$ 
of two fermions in terms of their s-wave scattering length $a_F$, 
into Eq. (\ref{law}) we exactly recover  
the Popov's 2D equation 
of state \cite{popov} of weakly-interacting bosons 
with scattering length $a_B$ \cite{castin2}, i.e. 
\beq 
n_B = {m_B\over 4\pi \hbar^2} \mu_B 
\ln{\left({4\hbar^2\over m_B \mu_B a_B^2 e^{2\gamma+1}}\right)} \; , 
\label{iopopov}
\eeq 
provided that we identify the effective bosonic scattering length $a_B$ with:
\beq 
a_B={1\over 2^{1/2}e^{1/4}} \ a_F = 0.551... a_F \: . 
\label{figata}
\eeq
With this choice of the relation between $a_B$ and $a_F$, the 
equation of state of  low density 
($n \ll m/(\pi \hbar^2) \epsilon_b$) fermions with a strong attractive 
interaction in 2D 
indeed coincides with the iterative Popov's 2D equation 
of state of weakly-interacting bosons \cite{popov,castin2}.
  
Remarkably, recent Monte Carlo results of 
Giorgini and Bertaina 
\cite{bertaina} give $a_B=0.55(4) a_F$, in full agreement with our 
determination, Eq. (\ref{figata}), and also with previous 
four-body scattering calculations of Petrov, Baranov 
and Shlyapnikov \cite{baranov}. Notice that, at the first 
iteration \cite{boronat} of Eq. (\ref{iopopov}), the leading 
contribution gives 
\beq 
\mu_B = {4\pi \hbar^2\over m_B} {n_B \over 
\ln{\left({1\over n_B a_B^2}\right)} } \; , 
\eeq
that is the equation of state found by Schick in 1971 \cite{schick}. 


In conclusion, by using a functional integral approach 
with dimensional regularization of Gaussian quantum fluctuations 
we have derived the flow equation 
of the running coupling constant which appears in the beyond-mean-field 
grand potential of the 2D attractive fermionic superfluid. 
We have found that in the BEC regime 
of the 2D BCS-BEC crossover this running coupling constant 
has a logarithmic dependence which exactly reproduces 
the Popov's equation of state of interacting 2D bosons. 
We have also shown that in the BCS regime quantum fluctuations 
are not divergent but they are however strongly suppressed. 
As a final comment, we
notice that our approach, limited to the 
quartic term in the low-momentum expansion of bosonic collective 
excitations,  cannot describe the entire the 2D BCS-BEC crossover.
In fact in the region where $\mu$ changes sign the coefficient 
$\lambda$  is extremely small indicating that further terms must be included 
in the theory. 

\section*{Acknowledgments}

The authors acknowledge for partial support Cariparo Foundation 
(Eccellenza project) and Ministero Istruzione 
Universita Ricerca (PRIN project). 
The authors thank Lara Benfatto, Gianluca Bertaina, Giacomo Bighin, 
Sergio Caprara, Luca Dell'Anna, 
Pieralberto Marchetti, Pierbiagio Pieri, Adriaan Schakel, and 
Jacques Tempere for enlightening discussions.


\begin{thebibliography}{99}

\bibitem{mermin} N.D. Mermin and H. Wagner, 
Phys. Rev. Lett. {\bf 17}, 133 (1966).

\bibitem{hohenberg} P.C. Hohenberg, Phys. Rev. {\bf 158}, 383 (1967).

\bibitem{coleman} S. Coleman, Commun. Math. Phys. {\bf 31}, 259 (1973).

\bibitem{nagaosa} N. Nagaosa, 
{\it Quantum Field Theory in Condensed Matter Physics} 
(Springer, Berlin, 1999). 

\bibitem{exp_MolecularBEC} M. Greiner, C. A. Regal,
and D. S. Jin, Nature {\bf 426}, 537 (2003); S. Jochim {\it et al.}
Science {\bf 302} (2003); M. W. Zwierlein {\it et al.}, Phys. Rev.
Lett. {\bf 91} 250401 (2003).

\bibitem{exp_Crossover} C. A. Regal, M. Greiner, and D. S. Jin, 
Phys. Rev. Lett.
{\bf 92}, 040403 (2004); M. W. Zwierlein {\it et al.}, Phys. Rev.
Lett. {\bf 92}, 120403 (2004);

\bibitem{chin} C. Chin {\it et al.}, Science {\bf 305}, 1128 (2004); 
Y. Shin et al., Nature (London) 451, 689 (2008).

\bibitem{loktev0} V.P. Gusynin, V.M. Loktev, and Sharapov, 
J. Exp. Theor. Phys. {\bf 88}, 685 (1999).

\bibitem{babaev} E. Babaev and H. Kleinert, 
Phys. Rev. B {\bf 59}, 12083 (1999). 

\bibitem{loktev} V.M. Loktev, R.M. Quick, and S.G. Sharapov, 
Phys. Rep. {\bf 349}, 1 (2001). 

\bibitem{demelo} S.S. Botelho and C.A.R. Sa de Melo, 
J. Low Temp. Phys. {\bf 140}, 409 (2005). 

\bibitem{tempere} J. Tempere, S.N. Klimin, and 
J.T. Devreese, Phys. Rev. A {\bf 79}, 053637 (2009).   

\bibitem{tempere2} S.N. Klimin, J.T. Devreese, 
and J. Tempere, New J. Phys. {\bf 14}, 103044 (2012). 

\bibitem{sala} L. Salasnich, P. Marchetti, and F. Toigo, 
Phys. Rev. A {\bf 88}, 053612 (2013). 

\bibitem{randeria2d} M. Randeria, J-M. Duan, and L-Y. Shieh, 
Phys. Rev. Lett. {\bf 62}, 981 (1989); M. Randeria, J-M. Duan, 
and L-Y. Shieh, Phys. Rev. B {\bf 41}, 327 (1990). 

\bibitem{bertaina} G. Bertaina and S. Giorgini, 
Phys. Rev. Lett. {\bf 106}, 110403 (2011); G. Bertaina, 
Eur. Phys. J. Special Topics {\bf 217}, 153 (2013).  

\bibitem{pieri} P. Pieri and G. Strinati, 
Phys. Rev. B {\bf 61}, 15370 (2000). 

\bibitem{hu} H.Hu, X.-J. Liu, and P. Drummond, Europhys. 
Lett. {\bf 74}, 574 (2006).

\bibitem{randeria3D} R.B. Diener, R. Sensarma, 
and M. Randeria, Phys. Rev. A {\bf 77}, 023626 (2008). 

\bibitem{dim-reg} G. 't Hooft and M. 
Veltman, Nucl. Phys. B {\bf 44}, 189 (1972). 

\bibitem{leibb} G. Leibbrandt, Rev. Mod. Phys. {\bf 47}, 849 (1975);  
see Eq. (4.24a). 

\bibitem{popov} V.N. Popov, Theor. Math. Phys. A {\bf 11}, 565 (1972). 

\bibitem{schick} M. Schick, Phys. Rev. A {\bf 3}, 1067 (1971). 

\bibitem{boronat} G.E. Astrakharchik, J. Boronat, J. Casulleras, 
I.L. Kurbakov, and Yu.E. Lozovik, Phys. Rev. A {\bf 79}, 
051602(R) (2009). 

\bibitem{baranov} D.S. Petrov, M.A. Baranov, and 
G.V. Shlyapnikov, Phys. Rev. A {\bf 67}, 031601 (2003). 

\bibitem{marini} M. Marini, F. Pistolesi, and G.C. Strinati, 
Eur. Phys. J B {\bf 1}, 151 (1998). 

\bibitem{castin1} C. Mora and Y. Castin, Phys. Rev. A
{\bf 67}, 053615 (2003).

\bibitem{schakel} A.M.J. Schakel, {\it Boulevard of Broken Symmetries}  
(World Scientific, Singapore, 2008).

\bibitem{marini2} M. Marini, M.Sc. thesis, 
University of Camerino (1998), unpublished.  

\bibitem{schakel0} A similar approach was developed in \cite{schakel}
for the study of zero-point energy of the weakly-interacting 2D Bose gas. 

\bibitem{remark} This result can be immediately obtained 
from Eqs. (10.12) and (10.29) of Ref. \cite{schakel}, 
which are valid in the mean-field BEC regime.  

\bibitem{kaku} M. Kaku, {\it Quantum Field Theory. A Modern Introduction} 
(Oxford, Univ. Press, 1993), cap. 14.5. 

\bibitem{castin2} C. Mora and Y. Castin, Phys. Rev. Lett. 
{\bf 102}, 180404 (2009). 


\end{thebibliography}
\end{document}